\documentclass[thmsa]{article}
\usepackage{amsmath}
\usepackage{sw20lart}
\usepackage{amsfonts}
\usepackage{amssymb}
\usepackage{graphicx}
\begin{document}
\noindent\textbf{{\LARGE {Multiple addition theorem for discrete }}}

\noindent\textbf{{\LARGE {and continuous nonlinear problems}}}

\begin{center}
\medskip%

\begin{tabular}
[c]{l}%
J. A. Zagrodzi\'{n}ski, T. Nikiciuk\\
{\footnotesize Institute of Physics, Polish Academy of Sciences, 02-668,
Warsaw, Poland}\\
{\footnotesize E-mail: zagro@ifpan.edu.pl}%
\end{tabular}

\bigskip
\end{center}

\noindent\textbf{Abstract}{\footnotesize .\ \ \ The addition relation for the
Riemann theta functions and for its limits, which lead to the appearance of
exponential functions in soliton type equations is discussed. The presented
form of addition property resolves itself to the factorization of }%
$N-${\footnotesize tuple product of the shifted functions and it seems to be
useful for analysis of soliton type continuous and discrete processes in the
}$N+1${\footnotesize \ space-time. A close relation with the natural
generalization of bi- and tri-linear operators into multiple linear operators
concludes the paper.}

\medskip


\section{Introduction}

The main goal of this note is the presentation of the role of the addition
property (AP), its relation to the famous bilinear operator formalism and its
universality, since using AP, the quasi-periodic processes and soliton ones
can be considered in an identical manner. An important step is however the
generalization of the leading to the bilinear operator standard version of the
AP to the version which can be linked with multilinear operators. It seems
that this generalized version can be useful in a case of multidimensional
soliton type problems.

There is opinion that a huge success of the bilinear formalism in the soliton
theory can be related to the AP for $\tau-$ functions which appear in the
majority of soliton equations%

\begin{equation}
\tau\left(  \mathbf{z}+\mathbf{w}\right)  \tau\left(  \mathbf{z}%
-\mathbf{w}\right)  =\sum_{\mathbf{\varepsilon}}W_{\mathbf{\varepsilon}%
}\left(  \mathbf{w}\right)  Z_{\varepsilon}\left(  \mathbf{z}\right)
,\label{c1}%
\end{equation}
where $\mathbf{z}=\mathbf{\kappa} x+\mathbf{\omega} t\in C^{g},$
$\mathbf{\varepsilon}\in Z_{2}^{g};\;\tau:C^{g}\rightarrow C$ and
$W_{\varepsilon}\left(  \mathbf{w}\right)  ,$ $Z_{\varepsilon}\left(
\mathbf{z}\right)  :C^{g}\times Z^{g}\rightarrow C.$

The essential point here is the factorization of the right hand side of
(\ref{c1}), in which\ functions $W_{\mathbf{\varepsilon}}$ and
$Z_{\mathbf{\varepsilon}}$ depend on $\mathbf{w}$ and on $\mathbf{z},$
respectively and exclusively. There are a few version of AP, according to
(\ref{c1}) scheme, \cite{BE}, \cite{Du}, \cite{Ig}, and the factorization
appears in each one. In applications to the soliton type equations, the
argument $\mathbf{z}$ usually depends on space and time, while $\mathbf{w}$
plays a role of a fixed constant parameter. In a few papers \cite{JZ1} it was
shown that (\ref{c1}) allows in a straightforward manner to determine
derivatives of logarithms of $\tau-$ function, which are useful in
differential version of soliton type equations. For the discrete soliton type
equations the form (\ref{c1}) has a direct and immediate application.

As it was shown in the cited references, a class containing exponential
functions as well as the Riemann theta functions has just the AP according to
(\ref{c1}).\ 

In order to illustrate an application of AP, we present below two examples:
the discrete Hirota equation and doubly discrete sine-Gordon equation
(dd-sGe). In the limit, when step tends to zero, the first - Hirota equation
has a trivial limit, while the second one (dd-sGe) becomes a standard sGe.

\bigskip

\section{The Hirota equation}

As an elementary example of the addition property we present the system of
dispersion equations for the functional (discrete) equations
\begin{align}
& a\tau\left(  x+h,y,t\right)  \tau\left(  x-h,y,t\right)  +b\tau\left(
x,y+h,t\right)  \tau\left(  x,y-h,t\right)  +c\tau\left(  x,y,t+h\right)
\tau\left(  x,y,t-h\right)
\begin{tabular}
[c]{l}%
=
\end{tabular}
\nonumber\\
& \;\;\;\;\;\;\;
\begin{tabular}
[c]{l}%
=
\end{tabular}
C\tau^{2}\left(  x,y,t\right)  ,\label{b1a}%
\end{align}
where $h$ is the step and $a,b,c,C-$are constant. When $C=0$ this equation is
known as the Hirota equation and then its soliton solutions one can be found
in \cite{Hi}. Quasiperiodic solutions are reported in \cite{KWZ} and another
class of solutions in \cite{Ve}. In the language of bilinear operator
(\ref{b1a}) can be written as
\begin{equation}
\left[  a\exp\left(  hD_{x}\right)  +b\exp\left(  hD_{y}\right)  +c\exp\left(
hD_{t}\right)  \right]  \left(  \tau\circ\tau\right)  =C\tau^{2},\label{b2}%
\end{equation}
where bilinear operator $D_{x}$ (in scalar version) is defined as%

\begin{equation}
\left(  D_{x}\right)  ^{n}\left(  \tau\circ\tau\right)  :=\left(  \partial
_{x}-\partial_{x_{1}}\right)  ^{n}\tau\left(  x,y,t\right)  \tau\left(
x_{1},y,t\right)  |_{x_{1}=x}=\left(  \partial_{s}\right)  ^{n}\left[
\tau\left(  x+s,y,t\right)  \tau\left(  x-s,y,t\right)  \right]
|_{s=0}.\label{b22}%
\end{equation}

Assuming that $\tau-$function argument is $\mathbf{z}=\mathbf{k}%
x+\mathbf{l}y+\mathbf{w}t$ $\in C^{g},$ equation (\ref{b1a}) can be rewritten as%

\begin{equation}
a\tau\left(  \mathbf{z+k}h\right)  \tau\left(  \mathbf{z-k}h\right)
+b\tau\left(  \mathbf{z+l}h\right)  \tau\left(  \mathbf{z-l}h\right)
+c\tau\left(  \mathbf{z+w}h\right)  \tau\left(  \mathbf{z-w}h\right)
=C\tau^{2}\left(  \mathbf{z}\right)  ,\label{b3}%
\end{equation}
which is just suitable for the application of the addition property. We obtain
functional equation%

\begin{equation}
\sum_{\varepsilon\in Z_{2}^{g}}\left[  aW_{\varepsilon}\left(  \mathbf{k}%
h\right)  +bW_{\varepsilon}\left(  \mathbf{l}h\right)  +cW_{\varepsilon
}\left(  \mathbf{w}h\right)  -CW_{\varepsilon}\left(  0\right)  \right]
Z_{\varepsilon}\left(  \mathbf{z}\right)  =0,\label{b4}%
\end{equation}
which in case of independent functions $Z_{\varepsilon}\left(  z\right)  $,
$\varepsilon\in Z_{2}^{g}$ leads to the system of algebraic equations%

\begin{equation}
aW_{\varepsilon}\left(  \mathbf{k}h\right)  +bW_{\varepsilon}\left(
\mathbf{l}h\right)  +cW_{\varepsilon}\left(  \mathbf{w}h\right)
-CW_{\varepsilon}\left(  0\right)  =0;\;\;\;\text{for\ each}%
\;\mathbf{\varepsilon}\in Z_{2}^{g}.\label{b5}%
\end{equation}
For the fixed type of solution (which determines the class of functions
$W_{\varepsilon}$ ) and for a fixed step $h,$ equations (\ref{b5}) determine
the relations between $a,b,c,C$ and $\mathbf{k,l,w}$. Thus the dispersion
equations (\ref{b5}) are valid both for soliton and for quasiperiodic
processes, and even for processes in the form of solitons on the periodic background.

\bigskip

\section{Discrete sine - Gordon equation.}

We consider the functional (discrete-discrete) version of sine - Gordon
equation (dd-sGe) in the form:%

\begin{equation}
\frac{1}{h^{2}}\sin\left(  \frac{u^{++}+u^{--}-u^{+-}-u^{-+}}{4}\right)
=\sin\left(  \frac{u^{++}+u^{--}+u^{+-}+u^{-+}}{4}\right)  ,\label{a1}%
\end{equation}
where $u^{\pm\pm}:=u\left(  \xi\pm h,\tau\pm h\right)  .$ It is obvious that
in the limit $h\rightarrow0$ equation (\ref{a1}) becomes the traditional sGe
in light cone coordinates%

\begin{equation}
u_{,\xi\tau}=\sin u.\label{a2}%
\end{equation}

We look for the quasi-periodic solutions (\ref{a1}) in the form, which is
identical (up to constant parameters) with solutions of (\ref{a2})%

\begin{equation}
u\left(  \xi,\tau\right)  =2i\ln\frac{\theta\left(  \mathbf{z+}\frac
{\mathbf{e}}{2}|B\right)  }{\theta\left(  \mathbf{z}|B\right)  },\label{a3}%
\end{equation}
where $\theta\left(  \mathbf{z}|B\right)  $ denotes the Riemann theta function
of argument $\mathbf{z=k}\xi\mathbf{+u}\tau\mathbf{\in C}^{g}$ and
parametrized by the Riemannian matrix $B\mathbf{\in C}^{g\times g}%
;\;\mathbf{e=[}1,...,1]\mathbf{\in} Z^{g},$ see e.g. \cite{BE}, \cite{JZ1}.
Moreover, in order to obtain the real solutions we require $\theta\left(
\mathbf{z+}\frac{\mathbf{e}}{2}|B\right)  =\left[  \theta\left(
\mathbf{z}|B\right)  \right]  ^{\ast}.$

Since
\begin{equation}
u\left(  \xi+h,\tau\pm h\right)  =2i\ln\frac{\theta\left(  \mathbf{z+}\left(
\mathbf{k\pm u}\right)  h\mathbf{+}\frac{\mathbf{e}}{2}|B\right)  }%
{\theta\left(  \mathbf{z+}\left(  \mathbf{k\pm u}\right)  h|B\right)  }%
=2i\ln\frac{\theta\left(  \mathbf{z+w}_{\pm}\mathbf{+}\frac{\mathbf{e}}%
{2}|B\right)  }{\theta\left(  \mathbf{z+w}_{\pm}|B\right)  },\label{a31}%
\end{equation}

and similar relations hold for $u\left(  \xi-h,\tau\pm h\right)  ,$ after
simple manipulations we can rewrite (\ref{a1}) in the form%

\begin{align}
& \frac{\theta\left(  \mathbf{z}+\mathbf{w}_{-}\right)  \theta\left(
\mathbf{z}-\mathbf{w}_{-}\right)  -h^{2}\theta\left(  \mathbf{z}%
+\mathbf{w}_{-}+\mathbf{e}/2\right)  \theta\left(  \mathbf{z}-\mathbf{w}%
_{-}+\mathbf{e}/2\right)  }{\theta\left(  \mathbf{z}+\mathbf{w}_{+}\right)
\theta\left(  \mathbf{z}-\mathbf{w}_{+}\right)  }
\begin{tabular}
[c]{l}%
=
\end{tabular}
\label{a4}\\
& \frac{\theta\left(  \mathbf{z}+\mathbf{w}_{-}+\mathbf{e}/2\right)
\theta\left(  \mathbf{z}-\mathbf{w}_{-}+\mathbf{e}/2\right)  -h^{2}%
\theta\left(  \mathbf{z}+\mathbf{w}_{-}\right)  \theta\left(  \mathbf{z}%
-\mathbf{w}_{-}\right)  }{\theta\left(  \mathbf{z}+\mathbf{w}_{+}%
+\mathbf{e}/2\right)  \theta\left(  \mathbf{z}-\mathbf{w}_{+}+\mathbf{e}%
/2\right)  },\nonumber
\end{align}
where \textbf{\ }$\mathbf{w}_{+}=\left(  \mathbf{k+u}\right)  h,\;\mathbf{w}%
_{-}=\left(  \mathbf{k-u}\right)  h.$\ Let us assume that they are constant
(with respect to $\xi$ and $\tau$). Both lead to the same result%

\begin{equation}
\theta\left(  \mathbf{z}+\mathbf{w}_{-}\right)  \theta\left(  \mathbf{z}%
-\mathbf{w}_{-}\right)  -h^{2}\theta\left(  \mathbf{z}+\mathbf{w}%
_{-}+\mathbf{e}/2\right)  \theta\left(  \mathbf{z}-\mathbf{w}_{-}%
+\mathbf{e}/2\right)  =\left(  C+1\right)  \theta\left(  \mathbf{z}%
+\mathbf{w}_{+}\right)  \theta\left(  \mathbf{z}-\mathbf{w}_{+}\right)
.\label{a5}%
\end{equation}

Now the addition property can be applied. The Riemann theta functions do have
the addition property :%

\begin{equation}
\theta\left(  \mathbf{z}+\mathbf{w|}B\right)  \theta\left(  \mathbf{z}%
-\mathbf{w|}B\right)  =\sum_{\varepsilon\in Z_{2}^{g}}W_{\varepsilon}\left(
\mathbf{w}\right)  \;\theta^{2}\left(  \mathbf{z+}\frac{\mathbf{\varepsilon}%
}{2}|B\right)  ,\label{a6}%
\end{equation}
for $\mathbf{z,}$ $\mathbf{w}$ $\in$ $C^{g}$. $W_{\varepsilon}$ coefficients
can be expressed by theta-constants \cite{JZ1}, but - and it is important - do
not depend on $\xi$ and $\tau.\;$Then equation (\ref{a5}) can be written as%

\begin{align}
& \sum_{\varepsilon\in Z_{2}^{g}}W_{\varepsilon}\left(  \mathbf{w}_{-}\right)
\theta^{2}\left(  \mathbf{z+}\frac{\mathbf{\varepsilon}}{2}|B\right)
-h^{2}\sum_{\varepsilon\in Z_{2}^{g}}W_{\varepsilon}\left(  \mathbf{w}%
_{-}\right)  \theta^{2}\left(  \mathbf{z+}\left(  \frac{\mathbf{e+\varepsilon
}}{2}\right)  |B\right)
\begin{tabular}
[c]{l}%
=
\end{tabular}
\label{a7}\\
& \;\;\;\;\;\;\;\;\;\;\;\;\;\;\;\;\;\;\;\;\;\;\;\;\;\;\;
\begin{tabular}
[c]{l}%
=
\end{tabular}
\left(  C+1\right)  \sum_{\varepsilon\in Z_{2}^{g}}W_{\varepsilon}\left(
\mathbf{w}_{+}\right)  \theta^{2}\left(  \mathbf{z+}\frac{\mathbf{\varepsilon
}}{2}|B\right)  ,\nonumber
\end{align}
or as a simple functional equation
\begin{equation}
\sum_{\varepsilon\in Z_{2}^{g}}\left[  W_{\varepsilon}\left(  \mathbf{w}%
_{-}\right)  -h^{2}W_{e-\varepsilon}\left(  \mathbf{w}_{-}\right)  -\left(
C+1\right)  W_{\varepsilon}\left(  \mathbf{w}_{+}\right)  \right]  \theta
^{2}\left(  \mathbf{z+}\frac{\varepsilon}{2}|B\right)  =0.\label{a8}%
\end{equation}

Since $\theta^{2}\left(  \mathbf{z+}\frac{\varepsilon}{2}|B\right)  $ labeled
by $\varepsilon\in Z_{2}^{g}$ form a set of linearly independent functions,
finally we arrive at the requirement that for any $\varepsilon\in Z_{2}^{g}$
\begin{equation}
W_{\varepsilon}\left(  \mathbf{w}_{-}\right)  -h^{2}W_{e-\varepsilon}\left(
\mathbf{w}_{-}\right)  -\left(  C+1\right)  W_{\varepsilon}\left(
\mathbf{w}_{+}\right)  =0.\label{a9}%
\end{equation}

Equations (\ref{a9}) determine of $\mathbf{k}$ and $\mathbf{u}$, (and $C$) and
these are linearly related to the propagation vectors $\mathbf{\kappa}$ and
angular frequencies $\mathbf{\omega}$ in laboratory coordinate system
($\mathbf{z}=\mathbf{\kappa} x+\mathbf{\omega} t).$ Therefore these equations
represent a system of dispersion equations for the discussed dd-sGe.
Nontrivial solutions of (\ref{a9}) determines the solutions of starting
equation (\ref{a1}), but for the higher $g,$ since then the system is
overdetermined, also some and even all elements of the matrix $B.$

As (dd-sGe)$\rightarrow$(sGe), with $h\rightarrow0,$ (\ref{a9}) also tends to
the dispersion equation for standard sGe%

\begin{equation}
\sum_{i,j}k_{i}u_{j}W_{\varepsilon,ij}+\frac{1}{2}\left(  \delta
_{e,\varepsilon}-c\delta_{\varepsilon,0}\right)  =0.\label{a10}%
\end{equation}
where $W_{\varepsilon,ij}:=$ $\frac{\partial^{2}}{\partial w_{i}\partial
w_{j}}W_{\varepsilon}\left(  \mathbf{w}\right)  |_{w=0}.$ In order to prove
this statement one can substitute a new constant $c=-C/h^{2}.$and use the
relations $W_{\varepsilon}\left(  \mathbf{z}\right)  =W_{\varepsilon}\left(
\mathbf{-z}\right)  $, $W_{\varepsilon}\left(  0\right)  =\delta
_{\varepsilon,0}$ , where $\delta$ represents the Kronecker symbol.

\bigskip

\section{Tri-linear operator}

In order to extend direct methods in the spirit of bilinear operator formalism
on the a broader class of equations, the trilinear operators $T$ and $T^{\ast
}$were introduced by R. Hirota, (c.f. \cite{GRH}, \cite{YTF})%

\begin{align}
\left(  T\right)  ^{n}\left(  \tau\circ\tau\circ\tau\right)   & :=\left(
\partial_{z}+j\partial_{w_{1}}+j^{2}\partial_{w_{2}}\right)  \tau\left(
z\right)  \tau\left(  w_{1}\right)  \tau\left(  w_{2}\right)  |_{w_{2}%
=w_{1}=z}\;,\label{d1}\\
\left(  T^{\ast}\right)  ^{n}\left(  \tau\circ\tau\circ\tau\right)   &
:=\left(  \partial_{z}+j^{2}\partial_{w_{1}}+j\partial_{w_{2}}\right)
\tau\left(  z\right)  \tau\left(  w_{1}\right)  \tau\left(  w_{2}\right)
|_{w_{2}=w_{1}=z}\;,\label{d2}%
\end{align}
where $j=\exp\left(  i2\pi/3\right)  .$

In this language e.g. the 5-th order equation of the Lax hierarchy
\begin{equation}
u_{5x}+10uu_{3x}+20u_{x}u_{xx}+30u^{2}u_{x}+u_{t}=0,\label{d3}%
\end{equation}
although lacking a bilinear representation, can be written in trilinear form \cite{GRH}%

\begin{equation}
\left(  7T_{x}^{6}+20T_{x}^{3}T_{x}^{\ast3}+27T_{x}T_{t}\right)  F\circ F\circ
F=0,\label{d4}%
\end{equation}
where $u=2\left(  \ln F\right)  _{xx}.$ We will return to this equation in the
last paragraph of this note. However, first let us try to generalize the
concept of bi- and tri-linear operators.

\bigskip

\section{Multiple addition property}

One can introduce multilinear (J-linear) operator by the relation%

\begin{equation}
\left(  T\right)  ^{n}\underset{J\;elements}{\underbrace{\left(  \tau
\circ....\circ\tau\right)  }}:=\left[  \left(  \partial_{z_{0}}+j\partial
_{z_{1}}+...+j^{J-1}\partial_{z_{J-1}}\right)  \prod_{i=0}^{J-1}\tau\left(
z_{i}\right)  \right]  _{z_{J-1}=...=z_{1}=z_{0}=z},\label{f1}%
\end{equation}
where $j=\exp\left(  i2\pi/J\right)  .$

We are convinced that the effectivity of the bi- and tri- linear operators
formalism depends on the relevant addition property of the functions to which
this formalism is applied. Therefore, the fundamental question is which class
of functions has the multiple AP. Instead of the class of exponential
functions appearing in the solutions of soliton type equations we focus our
attention on the Riemann theta functions constituting a more general class of
functions and expressing the quasi-periodic solutions. Obviously, exponential
functions can be considered as the limiting case of the theta functions.

Following \cite{BE}, \cite{Ig}, \cite{Du}, \cite{JZ1}, we adopt the definition
of the Riemann theta function as%

\begin{equation}
\theta\left(  \mathbf{z}|B\right)  =\sum_{n\in\mathbb{Z}^{g}}\exp\left[
i\pi\left(  2\left\langle \mathbf{z},\mathbf{n}\right\rangle +\left\langle
\mathbf{n},B\mathbf{n}\right\rangle \right)  \right]  ,\label{f2}%
\end{equation}
where $\mathbf{z}\in C^{g},\;B\in C^{g\times g}\;$is the Riemann matrix, (i.e.
symmetric with positively defined imaginary part) and $\left\langle
\mathbf{z},\mathbf{n}\right\rangle :=\sum_{j=1}^{g}z_{j}n_{j}.$

If $\;\mathbf{z,u}^{\left(  k\right)  }\in C^{g}${\Large ,}$\;k=0,..,J-1,$
and
\begin{equation}%
{\displaystyle\sum_{k=0}^{J-1}}
\mathbf{u}^{\left(  k\right)  }=0,\label{f2a}%
\end{equation}
one can prove \cite{JAZ} that%

\begin{align}
& \theta\left(  \mathbf{z}+\mathbf{u}^{\left(  0\right)  }|B\right)
\;\theta\left(  \mathbf{z}+\mathbf{u}^{\left(  1\right)  }|B\right)
\;\theta\left(  \mathbf{z}+\mathbf{u}^{\left(  2\right)  }|B\right)
...\theta\left(  \mathbf{z}+\mathbf{u}^{\left(  J-1\right)  }|B\right)
\begin{tabular}
[c]{l}%
=
\end{tabular}
\nonumber\\
& \;\;\;\;\;\;\;\;\;\;\;\;\;\;\;\;
\begin{tabular}
[c]{l}%
=
\end{tabular}
\sum_{\varepsilon\in\mathbb{Z}_{J}^{g}}W_{\varepsilon}\left(  \mathbf{u}%
^{\left(  0\right)  },...,\mathbf{u}^{\left(  J-1\right)  }\right)
\;Z_{\varepsilon}\left(  \mathbf{z}\right)  \;,\label{f3}%
\end{align}
where%

\begin{align}
& \;\;\;\;\;\;\;\;\;\;\;\;\;\;\;\;\;\;\;\;\;\;\;Z_{\varepsilon}\left(
\mathbf{z}\right)
\begin{tabular}
[c]{l}%
=
\end{tabular}
\exp\left[  i\pi\left(  2\left\langle \mathbf{z,\varepsilon}\right\rangle
+\left\langle \mathbf{\varepsilon},B\mathbf{\varepsilon}\right\rangle \right)
\right]  \;\theta\left(  J\mathbf{z}+B\mathbf{\varepsilon}|JB\right)
\label{f6}\\
& W_{\varepsilon}\left(  \mathbf{u}^{\left(  0\right)  },...,\mathbf{u}%
^{\left(  J-1\right)  }\right)
\begin{tabular}
[c]{l}%
=
\end{tabular}
\label{f7}\\
&
\begin{tabular}
[c]{l}%
=
\end{tabular}
\exp\left(  i2\pi<\mathbf{u}^{\left(  0\right)  },\mathbf{\varepsilon
>}\right)  \;\theta\left(
\begin{array}
[c]{c}%
\mathbf{u}^{\left(  0\right)  }-\mathbf{u}^{\left(  1\right)  }%
+B\mathbf{\varepsilon}\\
\mathbf{u}^{\left(  0\right)  }-\mathbf{u}^{\left(  2\right)  }%
+B\mathbf{\varepsilon}\\
...\\
\mathbf{u}^{\left(  0\right)  }-\mathbf{u}^{\left(  J-1\right)  }%
+B\mathbf{\varepsilon}%
\end{array}
\left|  \left[
\begin{array}
[c]{llll}%
2B & B & .. & B\\
B & 2B & .. & B\\
.. & .. & 2B & ..\\
B & B & .. & 2B
\end{array}
\right]  \right.  \right)  .\nonumber
\end{align}
All $\theta-$ functions are of order $g,$ except the appearing in (\ref{f7}),
which is of order $\left(  J-1\right)  g.$ The sum in (\ref{f3}) is over
$\varepsilon\in\mathbb{Z}_{J}^{g}\;i.e.$ over $g-$dimensional vectors whose
components are $0,1,...,J-1,$ and therefore the sum contains $J^{g}$ elements.
Equation (\ref{f3}), written here for theta functions, is a natural
generalization of AP (\ref{c1}). The same form has the generalized AP for
exponential functions defined by%

\begin{equation}
E\left(  \mathbf{z}|\tilde{B}\right)  =\sum_{n\in\mathbb{Z}_{2}^{g}}%
\exp\left[  i\pi\left(  2\left\langle \mathbf{z},\mathbf{n}\right\rangle
+\left\langle \mathbf{n},\tilde{B}\mathbf{n}\right\rangle \right)  \right]
,\label{p9}%
\end{equation}
which appear in the solutions of standard soliton equations.(Observe that the
difference between (\ref{f2}) and (\ref{p9}) is only in the number of elements
in the sum). It is convenient to assume that diagonal elements of \ matrix
$\tilde{B}\in C^{g\times g}$ are real. The constraint (\ref{f2a}) can be
eliminated easily by introducing new parameters \ $\mathbf{w}^{\left(
1\right)  },...,\mathbf{w}^{\left(  J-1\right)  }$ instead of $\mathbf{u}%
^{\left(  0\right)  },\mathbf{u}^{\left(  1\right)  }...,\mathbf{u}^{\left(
J-1\right)  }$ according to the relation%

\begin{equation}
\mathbf{w}^{\left(  k\right)  }=\frac{1}{j-1}\left(  \mathbf{u}^{\left(
k\right)  }-j\mathbf{u}^{\left(  k+1\right)  }\left(  1-\delta_{k,J-1}\right)
-j\mathbf{u}^{\left(  1\right)  }\delta_{k,J-1}\right)
,\;\;\;\;k=1,...,J-1,\label{p91}%
\end{equation}
where $\delta_{k,J-1}$ is the standard Kronecker symbol. Inversely
\begin{align}
\mathbf{u}^{\left(  k\right)  }  & =j^{1-k}\left(  j^{J-1}\sum_{m=1}%
^{k-1}j^{m}\mathbf{w}^{\left(  m\right)  }+\sum_{m=1}^{J-1}j^{m}%
\mathbf{w}^{\left(  m\right)  }\right)  \;\;\;\;k=1,...,J-1,\label{p92}\\
\mathbf{u}^{\left(  0\right)  }  & =\sum_{m=1}^{J-1}\mathbf{w}^{\left(
m\right)  }.\label{p93}%
\end{align}
In the Table 1 below we present relations between $\mathbf{u}^{\left(
p\right)  }$ and $\mathbf{w}^{\left(  q\right)  }$ for $J=2-5.$

\begin{center}%
\begin{tabular}
[c]{|lllll|}\hline
$J=$ & $\;\;2$ & $\;\;\;\;\;\;3$ & $\;\;\;\;\;\;\;\;\;\;4$ &
$\;\;\;\;\;\;\;\;\;\;\;\;\;\;5$\\\hline
$j=$ & $\;-1$ & $\;\;\exp\left(  i2\pi/3\right)  $ & $\;\;\;\;\;\;\;\;\;\;i$ &
$\;\;\;\;\;\;\;\;\;\exp\left(  i2\pi/5\right)  $\\\hline
$\mathbf{u}^{\left(  0\right)  }=$ & $\mathbf{w}^{\left(  1\right)  }$ &
$\mathbf{w}^{\left(  1\right)  }\;+\;\mathbf{w}^{\left(  2\right)  }\;$ &
$\mathbf{w}^{\left(  1\right)  }\;+\;\mathbf{w}^{\left(  2\right)
}\;+\;\mathbf{w}^{\left(  3\right)  }$ & $\mathbf{w}^{\left(  1\right)
}\;+\;\mathbf{w}^{\left(  2\right)  }\;+\;\mathbf{w}^{\left(  3\right)
}\mathbf{\;}+\;\mathbf{w}^{\left(  4\right)  }$\\
$\mathbf{u}^{\left(  1\right)  }=$ & $j\mathbf{w}^{\left(  1\right)  }$ &
$j\mathbf{w}^{\left(  1\right)  }+j^{2}\mathbf{w}^{\left(  2\right)  }$ &
$j\mathbf{w}^{\left(  1\right)  }+j^{2}\mathbf{w}^{\left(  2\right)  }%
+j^{3}\mathbf{w}^{\left(  3\right)  }$ & $j\mathbf{w}^{\left(  1\right)
}+j^{2}\mathbf{w}^{\left(  2\right)  }+j^{3}\mathbf{w}^{\left(  3\right)
}+j^{4}\mathbf{w}^{\left(  4\right)  }$\\
$\mathbf{u}^{\left(  2\right)  }=$ & \ \ $-$ & $j^{2}\mathbf{w}^{\left(
1\right)  }+j\mathbf{w}^{\left(  2\right)  }$ & $j^{3}\mathbf{w}^{\left(
1\right)  }+j\mathbf{w}^{\left(  2\right)  }+j^{2}\mathbf{w}^{\left(
3\right)  }$ & $j^{4}\mathbf{w}^{\left(  1\right)  }+j\mathbf{w}^{\left(
2\right)  }+j^{2}\mathbf{w}^{\left(  3\right)  }+j^{3}\mathbf{w}^{\left(
4\right)  }$\\
$\mathbf{u}^{\left(  3\right)  }=$ & \ \ $-$ & \ \ $\;\;\;\;\;-$ &
$j^{2}\mathbf{w}^{\left(  1\right)  }+j^{3}\mathbf{w}^{\left(  2\right)
}+j\mathbf{w}^{\left(  3\right)  }$ & $j^{3}\mathbf{w}^{\left(  1\right)
}+j^{4}\mathbf{w}^{\left(  2\right)  }+j\mathbf{w}^{\left(  3\right)  }%
+j^{2}\mathbf{w}^{\left(  4\right)  }$\\
$\mathbf{u}^{\left(  4\right)  }=$ & \ \ $-$ & \ \ $\;\;\;\;\;-$ &
\ \ $\;\;\;\;\;\;\;\;\;-$ & $j^{2}\mathbf{w}^{\left(  1\right)  }%
+j^{3}\mathbf{w}^{\left(  2\right)  }+j^{4}\mathbf{w}^{\left(  3\right)
}+j\mathbf{w}^{\left(  4\right)  }$\\\hline
\end{tabular}

Table 1.
\end{center}

\medskip

Note that the choice of $\mathbf{w}$ parameters is not unique. The set adopted
here, gives a full correspondence with trilinear operators introduced earlier
in soliton theory \cite{GRH}.\ Since for arbitrary integer, $J$ the sum
$\sum_{k=0}^{J-1}j^{k}=0,$ it is seen that the requirement (\ref{f2a}) is
fulfilled for any set $\mathbf{w}^{\left(  q\right)  }$.\ 

As it was already mentioned, there exist several versions of the addition
theorem for theta functions. To our knowledge only one form \cite{Ko},
\cite{BE} leads deals to the product of an arbitrary number of shifted theta
functions as in (\ref{f3}), but the r.h.s. is essentially different and
unusable for our purposes.

For the fixed $J$ (\ref{f3}) can be rewritten as
\begin{equation}
\exp\sum_{j=1}^{J}\left[  \ln\theta\left(  \mathbf{z}+\mathbf{u}^{\left(
j\right)  }\right)  -\ln\theta\left(  \mathbf{z}\right)  \right]  =\left[
\theta\left(  \mathbf{z}\right)  \right]  ^{-J}\sum_{\varepsilon
}W_{\mathbf{\varepsilon}}\left(  \mathbf{u}^{\left(  1\right)  }%
,...,\mathbf{u}^{\left(  J-1\right)  }\right)  Z_{\mathbf{\varepsilon}}\left(
\mathbf{z}\right)  \;.\label{p11}%
\end{equation}

Differentiating (\ref{p11}) with respect to different components of vectors
$\mathbf{u}^{\left(  k\right)  },$ ($k=1,...,J-1)$ we obtain%

\begin{align}
&  \frac{\partial^{p+...+q}}{\left(  \partial u_{\alpha}^{\left(  1\right)
}\right)  ^{p}...\left(  \partial u_{\beta}^{\left(  l\right)  }\right)  ^{q}%
}\exp\left[  \sum_{k=0}^{J-1}\ln\theta\left(  \mathbf{z}+\mathbf{u}^{\left(
k\right)  }\right)  -J\ln\theta\left(  \mathbf{z}\right)  \right]
\begin{tabular}
[c]{l}%
=
\end{tabular}
\label{p12}\\
&  =\left[  \theta\left(  \mathbf{z}\right)  \right]  ^{-J}\sum_{\varepsilon
}\left[  \frac{\partial^{p+...+q}}{\left(  \partial u_{\alpha}^{\left(
1\right)  }\right)  ^{p}...\left(  \partial u_{\beta}^{\left(  l\right)
}\right)  ^{q}}W_{\mathbf{\varepsilon}}\left(  \mathbf{u}^{\left(  1\right)
},...,\mathbf{u}^{\left(  J-1\right)  }\right)  \right]
Z_{\mathbf{\varepsilon}}\left(  \mathbf{z}\right)  ,\nonumber
\end{align}
where $j=\exp\left(  i2\pi/J\right)  \;\;$and $\mathbf{u}^{\left(  0\right)
}=\sum_{k=1}^{J-1}\mathbf{u}^{\left(  k\right)  }$ Changing now the
derivatives with respect $\mathbf{u}^{\left(  k\right)  }\;$on l.h.s. into
derivatives with respect to $\mathbf{z},$ we can find easily the relationship
between the derivatives of logarithms of theta functions ( with respect to
$\mathbf{z})$ and the derivatives of $W_{\mathbf{\varepsilon}}$ functions
(with respect to $\mathbf{u}^{\left(  k\right)  }).$ All relations become
simpler if $\mathbf{u}^{\left(  k\right)  }$ parameters are chosen to be zero.

As an example, the lowest nontrivial derivatives of $W_{\mathbf{\varepsilon}%
}\left(  \mathbf{u}^{\left(  1\right)  }\right)  $ and $W_{\mathbf{\varepsilon
}}\left(  \mathbf{u}^{\left(  1\right)  },\mathbf{u}^{\left(  2\right)
}\right)  ,\;$respectively, (at zero) up to fifth order are reported for
$J=2,$and $3$ below in Table 2 and Table 3. In the Appendix we report the
lowest nontrivial derivatives (up to 6-th order) also for $J=6.$

\medskip

\begin{center}%
\begin{tabular}
[c]{|llll|}\hline
&  &
\begin{tabular}
[c]{l}%
\ \\
{\large J=2}\\
\
\end{tabular}
& \\\hline
$\left(  W_{\mathbf{\varepsilon}}\right)  _{u_{\alpha}^{\left(  i\right)
}u_{\beta}^{\left(  i\right)  }}$ & $\Longleftrightarrow$ & $L_{\alpha\beta}$%
& \\
$\left(  W_{\mathbf{\varepsilon}}\right)  _{u_{\alpha}^{\left(  i\right)
}u_{\beta}^{\left(  i\right)  }u_{\gamma}^{\left(  i\right)  }u_{\delta
}^{\left(  i\right)  }}$ & $\Longleftrightarrow$ & $4\left(  3\times
L_{\alpha\underline{\beta}}L_{\underline{\gamma\delta}}+2L_{\alpha\beta
\gamma\delta}\right)  $ & \\
$\left(  W_{\mathbf{\varepsilon}}\right)  _{u_{\alpha}^{\left(  i\right)
}u_{\beta}^{\left(  i\right)  }u_{\gamma}^{\left(  i\right)  }u_{\delta
}^{\left(  i\right)  }u_{\zeta}^{\left(  i\right)  }u_{\mu}^{\left(  i\right)
}} $ & $\Longleftrightarrow$ & $8\left(  15\times L_{\alpha\underline{\beta}%
}L_{\underline{\gamma\delta}}L_{\underline{\zeta\mu}}\right)  +4\left(
15\times L_{\underline{\alpha\beta\gamma\delta}}L_{\underline{\zeta\mu}%
}\right)  +2L_{\alpha\beta\gamma\delta\zeta\mu}$ & \\\hline
\end{tabular}

Table 2.

\bigskip%

\begin{tabular}
[c]{|llll|}\hline
&  &
\begin{tabular}
[c]{l}%
\ \\
{\large J=3}\\
\
\end{tabular}
& \\\hline
$\left(  W_{\mathbf{\varepsilon}}\right)  _{u_{\alpha}^{\left(  i\right)
}u_{\beta}^{\left(  j\right)  }}$ & $\Longleftrightarrow$ & $2L_{\alpha\beta}$%
& $i\neq j$\\
$\left(  W_{\mathbf{\varepsilon}}\right)  _{u_{\alpha}^{\left(  i\right)
}u_{\beta}^{\left(  i\right)  }}$ & $\Longleftrightarrow$ & $L_{\alpha\beta}$%
& \\
$\left(  W_{\mathbf{\varepsilon}}\right)  _{u_{\alpha}^{\left(  i\right)
}u_{\beta}^{\left(  i\right)  }u_{\gamma}^{\left(  j\right)  }}$ &
$\Longleftrightarrow$ & $-L_{\alpha\beta\gamma}$ & $i\neq j$\\
$\left(  W_{\mathbf{\varepsilon}}\right)  _{u_{\alpha}^{\left(  i\right)
}u_{\beta}^{\left(  i\right)  }u_{\gamma}^{\left(  i\right)  }u_{\delta
}^{\left(  i\right)  }}$ & $\Longleftrightarrow$ & $4\left(  3\times
L_{\alpha\underline{\beta}}L_{\underline{\gamma\delta}}+2L_{\alpha\beta
\gamma\delta}\right)  $ & \\
$\left(  W_{\mathbf{\varepsilon}}\right)  _{u_{\alpha}^{\left(  i\right)
}u_{\beta}^{\left(  i\right)  }u_{\gamma}^{\left(  i\right)  }u_{\delta
}^{\left(  j\right)  }}$ & $\Longleftrightarrow$ & $2\left(  3\times
L_{\alpha\underline{\beta}}L_{\underline{\gamma\delta}}+2L_{\alpha\beta
\gamma\delta}\right)  $ & $i\neq j$\\
$\left(  W_{\mathbf{\varepsilon}}\right)  _{u_{\alpha}^{\left(  i\right)
}u_{\beta}^{\left(  i\right)  }u_{\gamma}^{\left(  j\right)  }u_{\delta
}^{\left(  j\right)  }}$ & $\Longleftrightarrow$ & $4L_{\alpha\beta}%
L_{\gamma\delta}+2\times L_{\alpha\underline{\gamma}}L_{\beta\underline
{\delta}}+L_{\alpha\beta\gamma\delta}$ & $i\neq j$\\
$\left(  W_{\mathbf{\varepsilon}}\right)  _{u_{\alpha}^{\left(  i\right)
}u_{\beta}^{\left(  i\right)  }u_{\gamma}^{\left(  i\right)  }u_{\delta
}^{\left(  i\right)  }u_{\zeta}^{\left(  j\right)  }}$ & $\Longleftrightarrow$%
& $-2\left(  6\times L_{\underline{\alpha\beta}}L_{\underline{\gamma\delta
}\zeta}\right)  -L_{\alpha\beta\gamma\delta\zeta}$ & $i\neq j$\\
$\left(  W_{\mathbf{\varepsilon}}\right)  _{u_{\alpha}^{\left(  i\right)
}u_{\beta}^{\left(  i\right)  }u_{\gamma}^{\left(  i\right)  }u_{\delta
}^{\left(  i\right)  }u_{\zeta}^{\left(  i\right)  }u_{\mu}^{\left(  i\right)
}} $ & $\Longleftrightarrow$ &
\begin{tabular}
[c]{l}%
$8\left(  15\times L_{\alpha\underline{\beta}}L_{\underline{\gamma\delta}%
}L_{\underline{\zeta\mu}}\right)  +4\left(  15\times L_{\underline{\alpha
\beta\gamma\delta}}L_{\underline{\zeta\mu}}\right)  $\\
$+2L_{\alpha\beta\gamma\delta\zeta\mu}$%
\end{tabular}
& $i\neq j$\\
$\left(  W_{\mathbf{\varepsilon}}\right)  _{u_{\alpha}^{\left(  i\right)
}u_{\beta}^{\left(  i\right)  }u_{\gamma}^{\left(  i\right)  }u_{\delta
}^{\left(  i\right)  }u_{\zeta}^{\left(  i\right)  }u_{\mu}^{\left(  j\right)
}} $ & $\Longleftrightarrow$ &
\begin{tabular}
[c]{l}%
$4\left(  15\times L_{\underline{\alpha\beta}}L_{\underline{\gamma\delta}%
}L_{\underline{\zeta}\mu}\right)  $\\
$+2\left(  5\times L_{\underline{\alpha\beta\gamma\delta}}L_{\underline{\zeta
}\mu}+10\times L_{\underline{\alpha\beta\gamma}\mu}L_{\underline{\delta\zeta}%
}\right)  +L_{\alpha\beta\gamma\delta\zeta\mu}$%
\end{tabular}
& $i\neq j$\\
$\left(  W_{\mathbf{\varepsilon}}\right)  _{u_{\alpha}^{\left(  i\right)
}u_{\beta}^{\left(  i\right)  }u_{\gamma}^{\left(  i\right)  }u_{\delta
}^{\left(  i\right)  }u_{\zeta}^{\left(  j\right)  }u_{\mu}^{\left(  j\right)
}} $ & $\Longleftrightarrow$ & $%
\begin{array}
[c]{l}%
8\left(  3\times L_{\alpha\underline{\beta}}L_{\underline{\gamma\delta}%
}L_{\zeta\mu}\right)  +12\left(  2\times L_{\alpha\beta}L_{\gamma
\underline{\zeta}}L_{\delta\underline{\mu}}\right)  +\\
4\times L_{\underline{\alpha\beta\gamma}\zeta}L_{\underline{\delta}\mu
}+4\times L_{\underline{\alpha\beta\gamma}\mu}L_{\underline{\delta}\zeta
}+4L_{\alpha\beta\gamma\delta}L_{\zeta\mu}+\\
2\left(  6\times L_{\underline{\alpha\beta}}L_{\underline{\gamma\delta}%
\zeta\mu}\right)  +L_{\alpha\beta\gamma\delta\zeta\mu}%
\end{array}
$ & $i\neq j$\\
$\left(  W_{\mathbf{\varepsilon}}\right)  _{u_{\alpha}^{\left(  i\right)
}u_{\beta}^{\left(  i\right)  }u_{\gamma}^{\left(  i\right)  }u_{\delta
}^{\left(  j\right)  }u_{\zeta}^{\left(  j\right)  }u_{\mu}^{\left(  j\right)
}} $ & $\Longleftrightarrow$ & $%
\begin{array}
[c]{l}%
4\left(  9\times L_{\alpha\beta}L_{\gamma\delta}L_{\zeta\mu}\right)  +6\times
L_{\alpha\underline{\underline{\delta}}}L_{\underline{\beta}\underline
{\underline{\xi}}}L_{\underline{\gamma}\underline{\underline{\mu}}}+\\
9\times L_{\underline{\alpha}\underline{\underline{\delta}}}L_{\underline
{\beta\gamma}\underline{\underline{\zeta\mu}}}+2\left(  6\times L_{\alpha
\underline{\beta}}L_{\underline{\gamma}\underline{\underline{\delta\zeta\mu}}%
}\right)  +9\times L_{\underline{\alpha\beta}\underline{\underline{\delta}}%
}L_{\underline{\gamma}\underline{\underline{\zeta\mu}}}\\
+L_{\alpha\beta\gamma\delta\zeta\mu}%
\end{array}
$ & $i\neq j$\\\hline
\end{tabular}

\medskip Table 3.
\end{center}

\medskip

The remaining derivatives of the order less than 6 - vanish. $L_{\alpha\beta
}:=$ $\partial_{z_{\alpha}z_{\beta}}\ln\theta\left(  \mathbf{z}|B\right)
|_{\mathbf{z}=0}$ etc., and we adopted here the shorthand notation which
includes all possible permutations with respect identically underlined
indices: e.g. $3\times L_{\alpha\underline{\beta}}L_{\underline{\gamma\delta}%
}:=L_{\alpha\beta}L_{\gamma\delta}+L_{\alpha\gamma}L_{\beta\delta}%
+L_{\alpha\delta}L_{\beta\gamma}.$

Some introductional applications of the above results can be found in Ref.
\cite{JAZ}.

\medskip

\section{Addition property versus multilinear operators}

The correspondence between the reported here system of dispersion equations
and bilinear operators is quite obvious. This affinity can be extended even
further. For fixed $J,$ let us introduce a hierarchy of operators $T^{\left(
n\right)  }$ labeled by $n=0,1,...,J-1$%

\begin{equation}
\left(  T^{\left(  n\right)  }\right)  ^{m}\left(  \circ\tau\right)  ^{\left(
J-1\right)  }:=\left(  T^{\left(  n\right)  }\right)  ^{m}\underset
{J-1}{\underbrace{\tau\circ\tau\circ...\circ\tau}}=\left(  T^{\left(
n\right)  }\right)  ^{m}\left[  \tau\left(  \mathbf{z+u}_{0}\right)
...\tau\left(  \mathbf{z+u}_{J-1}\right)  \right]  |_{\mathbf{u}%
_{0}=...\mathbf{u}_{J-1}=0}\;,\label{g1}%
\end{equation}

where%

\begin{align}
T^{\left(  0\right)  }  & =\sum_{m=0}^{J-1}\partial_{\mathbf{u}^{\left(
m\right)  }}\;,\label{g2}\\
T^{\left(  n\right)  }  & =\partial_{\mathbf{u}^{\left(  0\right)  }}%
+\sum_{m=J-n+1}^{J-1}j^{m}\partial_{\mathbf{u}^{\left(  m+n-J\right)  }}%
+\sum_{m=1}^{J-n}j^{m}\partial_{\mathbf{u}^{\left(  m+n-1\right)  }},\;\;0\neq
n<J\;,\label{g3}%
\end{align}
and differentiation relates of course to the indicated components of
$\mathbf{u}^{\left(  m\right)  }$ vectors i.e. $\mathbf{u}_{\alpha}^{\left(
m\right)  }.\;$For $J=2,3,4$ we have

\medskip

\begin{center}%
\begin{tabular}
[c]{|llll|}\hline
& $J=2$ & $J=3$ & $J=4$\\\hline
$T^{\left(  0\right)  }=$ & \multicolumn{1}{l|}{$\partial_{\mathbf{u}^{\left(
0\right)  }}+\partial_{\mathbf{u}^{\left(  1\right)  }}$} & $\partial
_{\mathbf{u}^{\left(  0\right)  }}+\;\partial_{\mathbf{u}^{\left(  1\right)
}}+\;\partial_{\mathbf{u}^{\left(  2\right)  }}$ & $\partial_{\mathbf{u}%
^{\left(  0\right)  }}+\;\partial_{\mathbf{u}^{\left(  1\right)  }}%
+\;\partial_{\mathbf{u}^{\left(  2\right)  }}+\;\partial_{\mathbf{u}^{\left(
3\right)  }}$\\
$T^{\left(  1\right)  }=$ & \multicolumn{1}{l|}{$\partial_{\mathbf{u}^{\left(
0\right)  }}+j\partial_{\mathbf{u}^{\left(  1\right)  }}$} & $\partial
_{\mathbf{u}^{\left(  0\right)  }}+j\partial_{\mathbf{u}^{\left(  1\right)  }%
}+j^{2}\partial_{\mathbf{u}^{\left(  2\right)  }}$ & $\partial_{\mathbf{u}%
^{\left(  0\right)  }}+j\partial_{\mathbf{u}^{\left(  1\right)  }}%
+j^{2}\partial_{\mathbf{u}^{\left(  2\right)  }}+j^{3}\partial_{\mathbf{u}%
^{\left(  3\right)  }}$\\
$T^{\left(  2\right)  }=$ & \multicolumn{1}{l|}{$\;\;\;\;\;-$} &
$\partial_{\mathbf{u}^{\left(  0\right)  }}+j^{2}\partial_{\mathbf{u}^{\left(
1\right)  }}+j\partial_{\mathbf{u}^{\left(  2\right)  }}$ & $\partial
_{\mathbf{u}^{\left(  0\right)  }}+j^{3}\partial_{\mathbf{u}^{\left(
1\right)  }}+j\partial_{\mathbf{u}^{\left(  2\right)  }}+j^{2}\partial
_{\mathbf{u}^{\left(  3\right)  }}$\\
$T^{\left(  3\right)  }=$ & \multicolumn{1}{l|}{$\;\;\;\;\;-$} &
$\;\;\;\;\;\;\;\;-$ & $\partial_{\mathbf{u}^{\left(  0\right)  }}%
+j^{2}\partial_{\mathbf{u}^{\left(  1\right)  }}+j^{3}\partial_{\mathbf{u}%
^{\left(  2\right)  }}+j\partial_{\mathbf{u}^{\left(  3\right)  }}.$\\\hline
\end{tabular}
\end{center}

\medskip

It is seen that for $J=2$ \ and $J=3\;$we have the standard two- and
tri-linear operators, respectively. However, for $J>3,$ the operator
$T^{\left(  2\right)  }\neq\left(  T^{\left(  1\right)  }\right)  ^{\ast}$
i.e. $T^{\left(  2\right)  }$ is not a complex conjugate to $T^{\left(
1\right)  }.$ For this reason the operators $T^{\left(  n\right)  }$
$(n=1,...,J-1)$ for fixed $J$, will called associated operators.\ Operator
$T^{\left(  0\right)  }$ is introduced here only for completeness.

\ \ \ \ \ Now, if $\tau-$ function from (\ref{g1}) has AP, the question arises
how it reflects on the $W_{\mathbf{\varepsilon}}\left(  \mathbf{w}^{\left(
1\right)  },...,\mathbf{w}^{\left(  J-1\right)  }\right)  $ functions?

Using (\ref{p92}) and (\ref{p93}) we have $\partial_{\mathbf{u}_{\alpha
}^{\left(  s\right)  }}/\partial_{\mathbf{w}_{\alpha}^{\left(  p\right)  }%
}=j^{1-s+p}\left(  1+\left(  j^{J-1}-1\right)  \delta_{0<p<s}\right)  \;\;$and therefore%

\begin{align}
\partial_{\mathbf{w}_{\alpha}^{\left(  p\right)  }}  & =\sum_{s=0}^{J-1}%
\frac{\partial_{\mathbf{u}_{\alpha}^{\left(  s\right)  }}}{\partial
_{\mathbf{w}_{\alpha}^{\left(  p\right)  }}}\partial_{\mathbf{u}_{\alpha
}^{\left(  s\right)  }}=\partial_{\mathbf{u}_{\alpha}^{\left(  0\right)  }%
}+j^{1+p}\sum_{s=0}^{J-1}\left[  j^{-s}\left(  1+\left(  j^{J-1}-1\right)
\delta_{0<p<s}\right)  \right]  \partial_{\mathbf{u}_{\alpha}^{\left(
s\right)  }}\nonumber\\
& =\partial_{\mathbf{u}_{\alpha}^{\left(  0\right)  }}+\sum_{m=J-n+1}%
^{J-1}j^{m}\partial_{\mathbf{u}_{\alpha}^{\left(  m+p-J\right)  }}+\sum
_{m=1}^{J-p}j^{m}\partial_{\mathbf{u}_{\alpha}^{\left(  m+p-1\right)  }%
}=T_{\alpha}^{\left(  p\right)  }\;.\label{g4}%
\end{align}

This means that if $\tau-$function has the AP, multilinear operators according
to (\ref{g1})-(\ref{g3}) reduce to the simple differentiation of the
$W_{\mathbf{\varepsilon}}\left(  \mathbf{w}^{\left(  1\right)  }%
,...,\mathbf{w}^{\left(  J-1\right)  }\right)  $ functions with respect to
their arguments. In the simplest cases of bi- and tri-linear operators this
assertion allows immediately to write the system of dispersion equation on the
basis of bi- or tri-linear approximation.

As an example, let us note the bi-linear form of Korteweg de Vries equation
and tri-linear form of the reduction of self-dual Yang Mills equation%

\begin{align}
\left(  D_{x}D_{t}+D_{x}^{4}\right)  \tau\circ\tau & =0,\label{g7}\\
\left(  T_{x}^{4}T_{z}^{\ast}+8T_{x}^{3}T_{z}T_{x}^{\ast}+9T_{x}^{2}%
T_{t}\right)  \tau\circ\tau\circ\tau & =0,\label{g8}%
\end{align}
coincide with the relevant dispersion equation systems
\begin{align}
\sum_{ij}\kappa_{i}\omega_{j}W_{\varepsilon,w_{i}w_{j}}+\sum_{ijkl}\kappa
_{i}\kappa_{j}\kappa_{k}\kappa_{l}W_{\varepsilon,w_{i}w_{j}w_{k}w_{l}}  &
=CW_{\varepsilon}\;,\label{g9}\\
\sum_{ijklm}\left(  \kappa_{i}\kappa_{j}\kappa_{k}\kappa_{l}\lambda
_{m}-8\kappa_{i}\kappa_{j}\kappa_{k}\lambda_{l}\kappa_{m}\right)
W_{\varepsilon,w_{i}w_{j}w_{k}w_{l}v_{m}}+9\sum_{ijk}\kappa_{i}\kappa
_{j}\omega_{k}W_{\varepsilon,w_{i}w_{j}w_{k}}  & =CW_{\varepsilon
}\;,\label{g10}%
\end{align}
where we assumed that the arguments of $\tau-$functions depend linearly on
space and time: $z_{i}=\kappa_{i}x+\omega_{i}t\;$and \ $z_{i}=\kappa
_{i}x+\lambda_{i}y+\omega_{i}t,$ respectively.\ Moreover, in the second
equation $W_{\varepsilon}=W_{\varepsilon}\left(  \mathbf{w}^{\left(  1\right)
},\mathbf{w}^{\left(  2\right)  }\right)  |_{w^{\left(  1\right)  }=w^{\left(
2\right)  }=0},$ depends on two vectors, designed for typographic reasons as
$w$ and $v.$ In both cases, $C$ appears as the integration constant and for
soliton solutions it vanishes, while for quasi-periodic solution it has to be
determined as an additional parameter. Finally equations (\ref{g9}) and
(\ref{g10}) should hold for any $\varepsilon\in\mathbb{Z}_{J}^{g},$ i.e. the
first one for $\varepsilon\in\mathbb{Z}_{2}^{g},$ and the second - for
$\varepsilon\in\mathbb{Z}_{3}^{g}.$

In conclusion, we expect that the reported here generalized addition property
can be useful for the analysis of multidimensional soliton equations.

The Authors acknowledge helpful discussions with Dr. M. Jaworski, Dr. R.
Pawlikowski and especially are grateful for remarks of Prof. S. Lewandowski.

\section{Appendix}

The lowest nontrivial derivatives (up to 6-th order) of of
$W_{\mathbf{\varepsilon}}\left(  \mathbf{w}^{\left(  1\right)  }%
,...,\mathbf{w}^{\left(  5\right)  }\right)  $ for $J=6.$

\begin{center}%
\begin{tabular}
[c]{|l|l|l|}\hline
$\left(  W_{\mathbf{\varepsilon}}\right)  _{u_{\alpha}^{\left(  i\right)
}u_{\beta}^{\left(  j\right)  }}$ & $2L_{\alpha\beta}$ & $i\neq j$\\\hline
$\left(  W_{\mathbf{\varepsilon}}\right)  _{u_{\alpha}^{\left(  i\right)
}u_{\beta}^{\left(  i\right)  }}$ & $L_{\alpha\beta}$ & \\\hline
$\left(  W_{\mathbf{\varepsilon}}\right)  _{u_{\alpha}^{\left(  i\right)
}u_{\beta}^{\left(  i\right)  }u_{\gamma}^{\left(  j\right)  }}$ &
$-L_{\alpha\beta\gamma}$ & $i\neq j$\\\hline
$\left(  W_{\mathbf{\varepsilon}}\right)  _{u_{\alpha}^{\left(  i\right)
}u_{\beta}^{\left(  j\right)  }u_{\gamma}^{\left(  k\right)  }}$ &
$-L_{\alpha\beta\gamma}$ & $i\neq j$\\\hline
$\left(  W_{\mathbf{\varepsilon}}\right)  _{u_{\alpha}^{\left(  i\right)
}u_{\beta}^{\left(  i\right)  }u_{\gamma}^{\left(  i\right)  }u_{\delta
}^{\left(  i\right)  }}$ & $4\left(  3\times L_{\alpha\underline{\beta}%
}L_{\underline{\gamma\delta}}+2L_{\alpha\beta\gamma\delta}\right)  $ &
\\\hline
$\left(  W_{\mathbf{\varepsilon}}\right)  _{u_{\alpha}^{\left(  i\right)
}u_{\beta}^{\left(  i\right)  }u_{\gamma}^{\left(  i\right)  }u_{\delta
}^{\left(  j\right)  }}$ & $2\left(  3\times L_{\alpha\underline{\beta}%
}L_{\underline{\gamma\delta}}+2L_{\alpha\beta\gamma\delta}\right)  $ & $i\neq
j$\\\hline
$\left(  W_{\mathbf{\varepsilon}}\right)  _{u_{\alpha}^{\left(  i\right)
}u_{\beta}^{\left(  i\right)  }u_{\gamma}^{\left(  j\right)  }u_{\delta
}^{\left(  j\right)  }}$ & $4L_{\alpha\beta}L_{\gamma\delta}+2\times
L_{\alpha\underline{\gamma}}L_{\beta\underline{\delta}}+L_{\alpha\beta
\gamma\delta}$ & $i\neq j$\\\hline
$\left(  W_{\mathbf{\varepsilon}}\right)  _{u_{\alpha}^{\left(  i\right)
}u_{\beta}^{\left(  i\right)  }u_{\gamma}^{\left(  j\right)  }u_{\delta
}^{\left(  k\right)  }}$ & $2L_{\alpha\beta}L_{\gamma\delta}+2\times
L_{\alpha\underline{\gamma}}L_{\beta\underline{\delta}}+L_{\alpha\beta
\gamma\delta}$ & $i,j,k-different$\\\hline
$\left(  W_{\mathbf{\varepsilon}}\right)  _{u_{\alpha}^{\left(  i\right)
}u_{\beta}^{\left(  j\right)  }u_{\gamma}^{\left(  k\right)  }u_{\delta
}^{\left(  l\right)  }}$ & $3\times L_{\alpha\underline{\beta}}L_{\underline
{\gamma\delta}}+L_{\alpha\beta\gamma\delta}$ & $i,j,k,l-different$\\\hline
$\left(  W_{\mathbf{\varepsilon}}\right)  _{u_{\alpha}^{\left(  i\right)
}u_{\beta}^{\left(  i\right)  }u_{\gamma}^{\left(  i\right)  }u_{\delta
}^{\left(  i\right)  }u_{\zeta}^{\left(  j\right)  }}$ & $-2\left(  6\times
L_{\underline{\alpha\beta}}L_{\underline{\gamma\delta}\zeta}\right)
-L_{\alpha\beta\gamma\delta\zeta}$ & $i\neq j$\\\hline
$\left(  W_{\mathbf{\varepsilon}}\right)  _{u_{\alpha}^{\left(  i\right)
}u_{\beta}^{\left(  i\right)  }u_{\gamma}^{\left(  i\right)  }u_{\delta
}^{\left(  j\right)  }u_{\zeta}^{\left(  k\right)  }}$ &
\begin{tabular}
[c]{l}%
$-2\left(  3\times L_{\underline{\alpha\beta}}L_{\underline{\gamma}\delta
\zeta}\right)  -3\times L_{\underline{\alpha}\delta}L_{\underline{\beta\gamma
}\zeta}$\\
$\;\;\;\;-3\times L_{\underline{\alpha}\zeta}L_{\underline{\beta\gamma}\delta
}-L_{\alpha\beta\gamma\delta\zeta}$%
\end{tabular}
& $j\neq i,$ $k\neq i$\\\hline
$\left(  W_{\mathbf{\varepsilon}}\right)  _{u_{\alpha}^{\left(  i\right)
}u_{\beta}^{\left(  i\right)  }u_{\gamma}^{\left(  j\right)  }u_{\delta
}^{\left(  j\right)  }u_{\zeta}^{\left(  k\right)  }}$ &
\begin{tabular}
[c]{l}%
$-2\left(  L_{\alpha\beta}L_{\gamma\delta\zeta}+L_{\gamma\delta}L_{\alpha
\beta\zeta}\right)  -2\times L_{\alpha\underline{\gamma}}L_{\beta
\underline{\delta}\zeta}$\\
$-2\times L_{\beta\underline{\gamma}}L_{\alpha\underline{\delta}\zeta}-4\times
L_{\underline{\alpha}\zeta}L_{\underline{\beta\delta\gamma}}-L_{\alpha
\beta\gamma\delta\zeta}$%
\end{tabular}
\ \ \  & $j\neq i,$ $k\neq i$\\\hline
$\left(  W_{\mathbf{\varepsilon}}\right)  _{u_{\alpha}^{\left(  i\right)
}u_{\beta}^{\left(  i\right)  }u_{\gamma}^{\left(  j\right)  }u_{\delta
}^{\left(  k\right)  }u_{\zeta}^{\left(  l\right)  }}$ &
\begin{tabular}
[c]{l}%
$+3\times\left(  L_{\alpha\underline{\gamma\delta}}L_{\beta\underline{\zeta}%
}+L_{\beta\underline{\gamma\delta}}L_{\alpha\underline{\zeta}}-L_{\alpha
\beta\underline{\gamma}}L_{\underline{\delta\zeta}}\right)  $\\
$\;\;\;\;\;\;\;\;+2L_{\alpha\beta}L_{\gamma\delta\zeta}-L_{\alpha\beta
\gamma\delta\zeta}$%
\end{tabular}
&
\begin{tabular}
[c]{l}%
$i,j,k,l-$\\
$different$%
\end{tabular}
\\\hline
$\left(  W_{\mathbf{\varepsilon}}\right)  _{u_{\alpha}^{\left(  i\right)
}u_{\beta}^{\left(  j\right)  }u_{\gamma}^{\left(  k\right)  }u_{\delta
}^{\left(  l\right)  }u_{\zeta}^{\left(  m\right)  }}$ & $-10\times
L_{\underline{\alpha\beta\gamma}}L_{\underline{\delta\zeta}}-L_{\alpha
\beta\gamma\delta\zeta}$ &
\begin{tabular}
[c]{l}%
$i,j,k,l,m-$\\
$different$%
\end{tabular}
\\\hline
\end{tabular}%

\begin{tabular}
[c]{|l|l|}\hline
$\left(  W_{\mathbf{\varepsilon}}\right)  _{u_{\alpha}^{\left(  i\right)
}u_{\beta}^{\left(  i\right)  }u_{\gamma}^{\left(  i\right)  }u_{\delta
}^{\left(  i\right)  }u_{\zeta}^{\left(  i\right)  }u_{\mu}^{\left(  i\right)
}} $ & $8\left(  15\times L_{\alpha\underline{\beta}}L_{\underline
{\gamma\delta}}L_{\underline{\zeta\mu}}\right)  +4\left(  15\times
L_{\underline{\alpha\beta\gamma\delta}}L_{\underline{\zeta\mu}}\right)
+2L_{\alpha\beta\gamma\delta\zeta\mu}$\\\hline
$\left(  W_{\mathbf{\varepsilon}}\right)  _{u_{\alpha}^{\left(  i\right)
}u_{\beta}^{\left(  i\right)  }u_{\gamma}^{\left(  i\right)  }u_{\delta
}^{\left(  i\right)  }u_{\zeta}^{\left(  i\right)  }u_{\mu}^{\left(  j\right)
}} $ &
\begin{tabular}
[c]{l}%
$4\left(  15\times L_{\underline{\alpha\beta}}L_{\underline{\gamma\delta}%
}L_{\underline{\zeta}\mu}\right)  +2\left(  5\times L_{\underline{\alpha
\beta\gamma\delta}}L_{\underline{\zeta}\mu}+10\times L_{\underline{\alpha
\beta\gamma}\mu}L_{\underline{\delta\zeta}}\right)  $\\
$+L_{\alpha\beta\gamma\delta\zeta\mu}$%
\end{tabular}
\\\hline
$\left(  W_{\mathbf{\varepsilon}}\right)  _{u_{\alpha}^{\left(  i\right)
}u_{\beta}^{\left(  i\right)  }u_{\gamma}^{\left(  i\right)  }u_{\delta
}^{\left(  i\right)  }u_{\zeta}^{\left(  j\right)  }u_{\mu}^{\left(  j\right)
}} $ & $%
\begin{array}
[c]{l}%
8\left(  3\times L_{\alpha\underline{\beta}}L_{\underline{\gamma\delta}%
}L_{\zeta\mu}\right)  +12\left(  2\times L_{\alpha\beta}L_{\gamma
\underline{\zeta}}L_{\delta\underline{\mu}}\right)  +4\times L_{\underline
{\alpha\beta\gamma}\zeta}L_{\underline{\delta}\mu}\\
+4\times L_{\underline{\alpha\beta\gamma}\mu}L_{\underline{\delta}\zeta
}+4L_{\alpha\beta\gamma\delta}L_{\zeta\mu}+2\left(  6\times L_{\underline
{\alpha\beta}}L_{\underline{\gamma\delta}\zeta\mu}\right)  +L_{\alpha
\beta\gamma\delta\zeta\mu}%
\end{array}
$\\\hline
$\left(  W_{\mathbf{\varepsilon}}\right)  _{u_{\alpha}^{\left(  i\right)
}u_{\beta}^{\left(  i\right)  }u_{\gamma}^{\left(  i\right)  }u_{\delta
}^{\left(  j\right)  }u_{\zeta}^{\left(  j\right)  }u_{\mu}^{\left(  j\right)
}} $ & $%
\begin{array}
[c]{l}%
4\left(  9\times L_{\underline{\alpha\beta}}L_{\underline{\gamma}%
\underline{\underline{\delta}}}L_{\underline{\underline{\zeta\mu}}}\right)
+6\times L_{\alpha\underline{\underline{\delta}}}L_{\underline{\beta
}\underline{\underline{\xi}}}L_{\underline{\gamma}\underline{\underline{\mu}}%
}+9\times L_{\underline{\alpha}\underline{\underline{\delta}}}L_{\underline
{\beta\gamma}\underline{\underline{\zeta\mu}}}\\
+2\left(  3\times L_{\underline{\alpha\beta}}L_{\underline{\gamma}\delta
\zeta\mu}+3\times L_{\alpha\beta\gamma\underline{\delta}}L_{\underline
{\zeta\mu}}\right)  +9\times L_{\underline{\alpha\beta}\underline
{\underline{\delta}}}L_{\underline{\gamma}\underline{\underline{\zeta\mu}}%
}+L_{\alpha\beta\gamma\delta\zeta\mu}%
\end{array}
$\\\hline
$\left(  W_{\mathbf{\varepsilon}}\right)  _{u_{\alpha}^{\left(  i\right)
}u_{\beta}^{\left(  i\right)  }u_{\gamma}^{\left(  i\right)  }u_{\delta
}^{\left(  i\right)  }u_{\zeta}^{\left(  j\right)  }u_{\mu}^{\left(  k\right)
}} $ & $%
\begin{array}
[c]{l}%
4\left(  3\times L_{\alpha\underline{\beta}}L_{\underline{\gamma\delta}%
}L_{\zeta\mu}\right)  +2\left(  12\times L_{\underline{\alpha\beta}%
}L_{\underline{\gamma}\underline{\underline{\zeta}}}L_{\underline{\delta
}\underline{\underline{\mu}}}\right)  +8\times L_{\underline{\alpha}%
\underline{\underline{\zeta}}}L_{\underline{\beta\gamma\delta}\underline
{\underline{\mu}}}+\\
2\left(  6\times L_{\underline{\alpha\beta}}L_{\underline{\gamma\delta}%
\zeta\mu}\right)  +2L_{\alpha\beta\gamma\delta}L_{\zeta\mu}+6\times
L_{\underline{\alpha\beta}\zeta}L_{\underline{\gamma\delta}\mu}+L_{\alpha
\beta\gamma\delta\zeta\mu}%
\end{array}
$\\\hline
$\left(  W_{\mathbf{\varepsilon}}\right)  _{u_{\alpha}^{\left(  i\right)
}u_{\beta}^{\left(  i\right)  }u_{\gamma}^{\left(  i\right)  }u_{\delta
}^{\left(  j\right)  }u_{\zeta}^{\left(  j\right)  }u_{\mu}^{\left(  k\right)
}} $ & $%
\begin{array}
[c]{l}%
6\times L_{\alpha\underline{\underline{\delta}}}L_{\underline{\beta}%
\underline{\underline{\zeta}}}L_{\underline{\gamma}\mu}+4\left(  3\times
L_{\underline{\alpha\beta}}L_{\delta\zeta}L_{\underline{\gamma}\mu}\right)
+2\left(  6\times L_{\underline{\alpha\beta}}L_{\underline{\gamma}%
\underline{\underline{\delta}}}L_{\underline{\underline{\zeta}}\mu}\right) \\
+3\times L_{\underline{\alpha\beta}\delta\zeta}L_{\underline{\gamma}\mu
}+6\times L_{\underline{\alpha\beta}\underline{\underline{\delta}}\mu
}L_{\underline{\gamma}\underline{\underline{\zeta}}}+2L_{\gamma\delta
}L_{\alpha\beta\zeta\mu}+2\times L_{\alpha\beta\gamma\underline{\delta}%
}L_{\underline{\zeta}\mu}\\
+2\left(  3\times L_{\underline{\alpha\beta}}L_{\underline{\gamma}\delta
\zeta\mu}\right)  +6\times L_{\underline{\alpha\beta}\underline{\underline
{\delta}}}L_{\underline{\gamma}\underline{\underline{\zeta}}\mu}+3\times
L_{\underline{\alpha}\underline{\beta}\mu}L_{\underline{\gamma}\delta\zeta
}+L_{\alpha\beta\gamma\delta\zeta\mu}%
\end{array}
$\\\hline
$\left(  W_{\mathbf{\varepsilon}}\right)  _{u_{\alpha}^{\left(  i\right)
}u_{\beta}^{\left(  i\right)  }u_{\gamma}^{\left(  i\right)  }u_{\delta
}^{\left(  j\right)  }u_{\zeta}^{\left(  k\right)  }u_{\mu}^{\left(  l\right)
}} $ & $%
\begin{array}
[c]{l}%
6\times L_{\underline{\alpha}\underline{\underline{\delta}}}L_{\underline
{\beta}\underline{\underline{\zeta}}}L_{\underline{\gamma}\underline
{\underline{\mu}}}+2\left(  9\times L_{\underline{\alpha\beta}}L_{\underline
{\gamma}\underline{\underline{\delta}}}L_{\underline{\underline{\zeta\mu}}%
}\right)  +3\times L_{\alpha\beta\gamma\underline{\delta}}L_{\underline
{\zeta\mu}}\\
+9\times L_{\underline{\alpha\beta}\underline{\underline{\delta\zeta}}%
}L_{\underline{\gamma}\underline{\underline{\mu}}}+2\left(  3\times
L_{\underline{\alpha\beta}}L_{\underline{\gamma}\delta\zeta\mu}\right)
+8\times L_{\underline{\alpha\beta}\underline{\underline{\delta}}%
}L_{\underline{\gamma}\underline{\underline{\zeta\mu}}}+L_{\alpha\beta
\gamma\delta\zeta\mu}%
\end{array}
$\\\hline
$\left(  W_{\mathbf{\varepsilon}}\right)  _{u_{\alpha}^{\left(  i\right)
}u_{\beta}^{\left(  i\right)  }u_{\gamma}^{\left(  j\right)  }u_{\delta
}^{\left(  j\right)  }u_{\zeta}^{\left(  k\right)  }u_{\mu}^{\left(  k\right)
}} $ & $%
\begin{array}
[c]{l}%
8L_{\alpha\beta}L_{\gamma\delta}L_{\zeta\mu}+2\times\left(  L_{\underline
{\alpha}\gamma}L_{\underline{\beta}\delta}L_{\zeta\mu}+L_{\underline{\zeta
}\alpha}L_{\underline{\mu}\beta}L_{\gamma\delta}+L_{\underline{\gamma}\zeta
}L_{\underline{\delta}\mu}L_{\alpha\beta}\right) \\
8\times L_{\underline{\alpha}\underline{\underline{\underline{\zeta}}}%
}L_{\underline{\beta}\underline{\underline{\gamma}}}L_{\underline
{\underline{\delta}}\underline{\underline{\underline{\mu}}}}+6\times
L_{\underline{\alpha\beta}\underline{\underline{\gamma}}}L_{\underline
{\underline{\delta}}\underline{\underline{\underline{\zeta\mu}}}}+2\left(
3\times L_{\underline{\alpha\beta}\underline{\underline{\gamma\delta}}%
}L_{\underline{\underline{\underline{\zeta\mu}}}}\right)  +\\
12\times L_{\underline{\alpha}\underline{\beta}\underline{\underline{\gamma}%
}\underline{\underline{\underline{\zeta}}}}L_{\underline{\underline{\delta}%
}\underline{\underline{\underline{\mu}}}}+4\times L_{\underline{\alpha
}\underline{\underline{\gamma}}\zeta}L_{\underline{\beta}\underline
{\underline{\delta}}\mu}+L_{\alpha\beta\gamma\delta\zeta\mu}%
\end{array}
$\\\hline
$\left(  W_{\mathbf{\varepsilon}}\right)  _{u_{\alpha}^{\left(  i\right)
}u_{\beta}^{\left(  i\right)  }u_{\gamma}^{\left(  j\right)  }u_{\delta
}^{\left(  j\right)  }u_{\zeta}^{\left(  k\right)  }u_{\mu}^{\left(  l\right)
}} $ & $%
\begin{array}
[c]{l}%
4L_{\alpha\beta}L_{\gamma\delta}L_{\zeta\mu}+2\left(  4\times L_{\underline
{\alpha\beta}}L_{\underline{\underline{\gamma}}\zeta}L_{\underline
{\underline{\delta}}\mu}\right)  +10\times L_{\underline{\alpha}%
\underline{\underline{\underline{\zeta}}}}L_{\underline{\beta}\underline
{\underline{\gamma}}}L_{\underline{\underline{\delta}}\underline
{\underline{\underline{\mu}}}}\\
+2\left(  2\times L_{\underline{\alpha\beta}}L_{\underline{\underline
{\gamma\delta}}\zeta\mu}\right)  +13\times L_{\underline{\alpha}%
\underline{\beta}\underline{\underline{\gamma}}\underline{\underline
{\underline{\zeta}}}}L_{\underline{\underline{\delta}}\underline
{\underline{\underline{\mu}}}}+6\times L_{\underline{\alpha\beta}%
\underline{\underline{\gamma}}}L_{\underline{\underline{\delta}}%
\underline{\underline{\underline{\zeta}}}\underline{\underline{\underline{\mu
}}}}+\\
4\times L_{\alpha\underline{\gamma}\underline{\underline{\zeta}}}%
L_{\beta\underline{\delta}\underline{\underline{\mu}}}+L_{\alpha\beta
\gamma\delta\zeta\mu}%
\end{array}
$\\\hline
$\left(  W_{\mathbf{\varepsilon}}\right)  _{u_{\alpha}^{\left(  i\right)
}u_{\beta}^{\left(  i\right)  }u_{\gamma}^{\left(  j\right)  }u_{\delta
}^{\left(  k\right)  }u_{\zeta}^{\left(  l\right)  }u_{\mu}^{\left(  m\right)
}} $ & $%
\begin{array}
[c]{l}%
2\left(  3\times L_{\alpha\beta}L_{\gamma\underline{\delta}}L_{\underline
{\zeta}\underline{\mu}}\right)  +12\times L_{\underline{\alpha}\underline
{\underline{\gamma}}}L_{\underline{\beta}\underline{\underline{\delta}}%
}L_{\underline{\underline{\zeta}}\underline{\underline{\mu}}}+6\times
L_{\alpha\beta\underline{\gamma}\underline{\delta}}L_{\underline{\zeta
}\underline{\mu}}\\
+2L_{\alpha\beta}L_{\gamma\delta\zeta\mu}+8\times L_{\underline{\alpha
}\underline{\underline{\gamma}}}L_{\underline{\beta}\underline{\underline
{\delta}}\underline{\underline{\zeta}}\underline{\underline{\mu}}}+\\
4\times L_{\alpha\beta\underline{\gamma}}L_{\underline{\delta}\underline
{\zeta}\underline{\mu}}+4\times L_{\alpha\underline{\gamma}\underline{\delta}%
}L_{\beta\underline{\zeta}\underline{\mu}}+L_{\alpha\beta\gamma\delta\zeta\mu}%
\end{array}
$\\\hline
\end{tabular}
\end{center}
\end{document}